\documentclass[a4paper]{article}

\usepackage{epsfig}
\usepackage{hyperref}
\usepackage{a4}
\usepackage{latexsym, amssymb} 
\usepackage{amsmath}
\usepackage{comment}
\usepackage{cite}
\usepackage{comment}

\usepackage{datetime}

\usepackage{booktabs}
\usepackage{multirow}

\newcommand{\lapprox}{%
\mathrel{%
\setbox0=\hbox{$<$}
\raise0.6ex\copy0\kern-\wd0
\lower0.65ex\hbox{$\sim$}
}}
\newcommand{\gapprox}{%
\mathrel{%
\setbox0=\hbox{$>$}
\raise0.6ex\copy0\kern-\wd0
\lower0.65ex\hbox{$\sim$}
}}

\catcode`@=11 
\def \gsim{\mathrel{\mathpalette\@versim>}}
\def \lsim{\mathrel{\mathpalette\@versim<}}
\def \@versim#1#2{\lower0.4ex\vbox{\baselineskip\z@skip\lineskip\z@skip
     \lineskiplimit\z@\ialign{$\m@th#1\hfil##\hfil$%
     \crcr#2\crcr\sim\crcr}}}
\catcode`@=12 

\makeatletter
 
 \@addtoreset{equation}{section}
\makeatother

\begin{document}

\hspace*{10cm}

\begin{center}

{\Large\bf A set of new observables in the process $e^+_{}e^-_{}\to ZHH$.}\\[15mm]

{\bf Junya Nakamura}
\\                    
\vskip .1cm
{Institut f\"ur theoretische Physik, Universit\"at T\"ubingen, Germany} \\
\vskip .2cm
{\small junya.nakamura@itp.uni-tuebingen.de} \\

\end{center}

\vskip 1.5cm

\begin{abstract}

Consequences of non-standard Higgs couplings in the final-state distributions of the process $e^+_{}e^-_{} \to ZHH$ are studied. We derive an analytic expression for the differential cross section, which has in the most general case 9 non-zero functions. These functions are the coefficients of 9 angular terms, depend on the Higgs couplings, and can be experimentally measured as observables. Symmetry properties of these 9 functions are carefully discussed, and they are divided into 4 categories under CP and $\mathrm{CP\widetilde{T}}$. The relations between our observables and the observables which exist in the literature are also clarified. We numerically study the dependence of our observables on the parameters in an effective Lagrangian for the Higgs couplings. It is shown that these new observables depend on most of the effective Lagrangian parameters in different ways from the total cross section. A benefit from longitudinally polarized $e^+_{}e^-_{}$ beams is also discussed.

\end{abstract}



\section{Introduction}\label{sec:intro}

One of the main targets of experiments at future $e^+_{}e^-_{}$ colliders is the measurement of the trilinear self-coupling $\lambda_H^{}$ of the Higgs boson~\cite{Baer:2013cma, Asner:2013psa, Moortgat-Picka:2015yla, Fujii:2015jha, 2012arXiv1202.5940L}. 
The process $e^+_{}e^-_{}\to ZHH$ is expected to be the best reaction to measure $\lambda_H^{}$ in the earlier stage of experiments~\cite{Ilyin:1995iy, Boudjema:1995cb, Miller:1999ct, Battaglia:2001nn, Castanier:2001sf, Yasui:2002se, Takubo:2009rn, Giannelli:2009fe, Baur:2009uw, Tian:2013qmi, Strube:2016eje} (i.e. the center-of-mass energy $\sqrt{s}\simeq 500$ GeV) for the discovered Higgs boson with mass $\simeq 125$ GeV~\cite{Aad:2012tfa, Chatrchyan:2012xdj}. The process is sensitive to the couplings $HHZZ$~\cite{Gounaris:1979px} and $HHZ\gamma$, too, which cannot be accessed through single Higgs boson production processes such as $e^+_{}e^-_{}\to ZH$. \\

Due to its importance, many authors have investigated the process. The total cross section in the standard model (SM) was calculated for the first time in ref.~\cite{Gounaris:1979px}. This work was followed by several studies~\cite{Barger:1988jk, Dai:1989tq}. These papers numerically calculated various distributions of the final particles, too. 
The one-loop radiative corrections to the process were calculated in refs.~\cite{Zhang:2003jy, Belanger:2003ya}. 
The total cross section in the minimal supersymmetric extension of the SM~\cite{Kamoshita:1996xv, Djouadi:1996ah, Osland:1998hv, Djouadi:1999gv, Djouadi:2008gy}, that in composite Higgs models~\cite{Grober:2013fpx, Kanemura:2016tan} and that in other several new physics models~\cite{Asakawa:2010xj} have also been studied in detail. 
Refs.~\cite{Djouadi:1996ah, Osland:1998hv, Djouadi:1999gv} included the analytic form of the 2 Higgs energy distributions. 
The accuracy of measuring $\lambda_H^{}$ through the process $e^+_{}e^-_{}\to ZHH$ at future $e^+_{}e^-_{}$ colliders has been studied in refs.~\cite{Ilyin:1995iy, Boudjema:1995cb, Miller:1999ct, Battaglia:2001nn, Castanier:2001sf, Yasui:2002se, Takubo:2009rn, Giannelli:2009fe, Baur:2009uw, Tian:2013qmi, Strube:2016eje} by assuming that all the other couplings are the SM values. The expected constrains on several parameters (including parameters which affect $\lambda_H^{}$) in an effective Lagrangian have been discussed in refs.~\cite{Barger:2003rs, Contino:2013gna, Kumar:2014zra, Barklow:2017awn}. Ref.~\cite{Contino:2013gna} included the analytic form for the invariant mass distribution of the 2 Higgs bosons. \\

Most of the above studies, however, restricted themselves to the total cross section as inputs from the experiments. This will not be a problem, if one intends to determine only one parameter such as $\lambda_H^{}$. However, if one intends to determine more than one parameters at the same time as studied in refs.~\cite{Barger:2003rs, Contino:2013gna, Kumar:2014zra, Barklow:2017awn}, measuring only the total cross section is not enough and one needs to consider other observables such as the invariant mass distribution of the 2 Higgs bosons~\cite{Contino:2013gna, Kumar:2014zra}. 
The purpose of this paper is to introduce such observables in a rather different way. We find 9 observables as the coefficients of 9 angular terms in the differential cross section, one of which is directly related to the total cross section. The other 8 observables have not been studied in the literature. 
Symmetry properties of the 9 observables are clarified, and they are divided into 4 categories under CP and $\mathrm{CP\widetilde{T}}$~\cite{Hagiwara:1986vm}: 4 even-even, 1 even-odd, 2 odd-even and 2 odd-odd. The CP-odd observables directly measure CP non-conservation and the $\mathrm{CP\widetilde{T}}$-odd observables re-scattering effects. To our knowledge, any CP-odd and/or $\mathrm{CP\widetilde{T}}$-odd observables in this process have not been constructed so far. \\

This paper is organized as follows. 
In Section~\ref{kinematics}, we explain kinematics of the process $e^+_{}e^-_{}\to ZHH$. 
In Section~\ref{observables}, we give an analytic expression for the differential cross section. The differential cross section has 9 non-zero functions in the most general case and these 9 functions can be measured experimentally. 
We re-derive the analytic forms of the observables which exist in the literature and have been widely used, such as the invariant mass distribution of the 2 Higgs bosons. 
We show that all of these observables are directly related to just one of our 9 functions. 
The analytic form of the $Z$ boson polar angle distribution is also derived. 
In Section~\ref{symmetry}, the symmetry properties of the 9 functions are studied. 
In Section~\ref{numerical}, we form observables in terms of our 9 functions and numerically study the dependence of these observables on the parameters in an effective Lagrangian for the Higgs couplings.
We show that our new observables depend on most of the effective Lagrangian parameters in different ways than the total cross section. It is also shown that the sensitivity of the $\mathrm{CP\widetilde{T}}$-odd observable can be significantly enhanced by means of longitudinally polarized $e^+_{}e^-_{}$ beams. 
Section~\ref{summary} summarizes our findings.

\section{Kinematics}\label{kinematics}

\begin{figure}[t]
\centering
\includegraphics[scale=0.6]{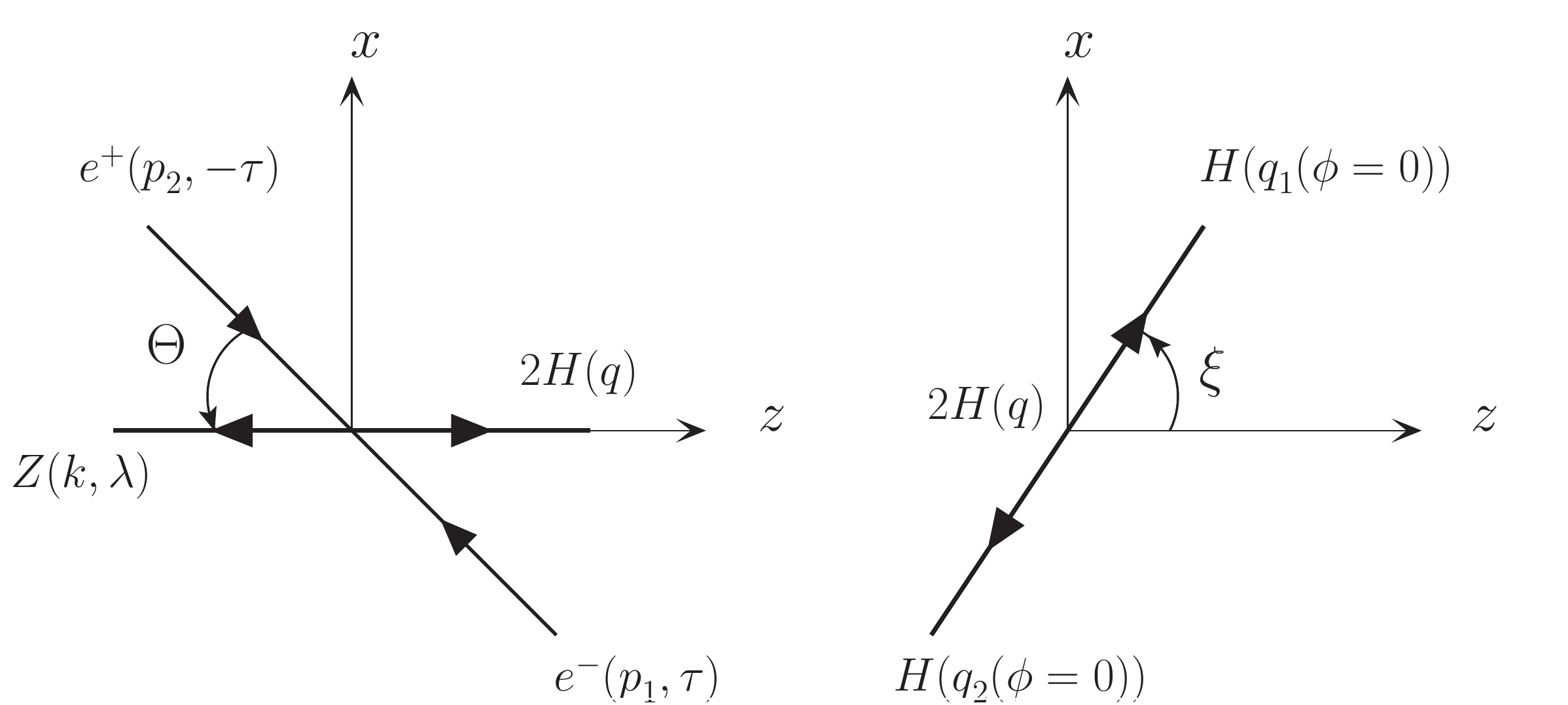}
\caption{\small {\it Left}: The coordinate system in the c.m. frame of the colliding $e^+_{}e^-_{}$ beams. The four-momentum and helicity of each particle are shown in parenthesis. A single object as the sum of the two Higgs bosons is represented by $2H$, whose four-momentum is $q^{\mu}_{}$. The direction of $\vec{q}$ is chosen as the $z$-axis and the $\vec{p}_1^{} \times \vec{k}$ direction as the $y$-axis.
{\it Right}: The coordinate system in the c.m. frame of the two Higgs bosons (i.e. the rest frame of $q^{\mu}_{}$). The parametrization of the four-momenta in this frame is given in eq.~(\ref{eq:momentumQq1q2}).}
\label{figure:ee-ZHH-frames}
\end{figure}

Our coordinate system in the center-of-mass (c.m.) frame of the colliding $e^+_{}e^-_{}$ beams is described in Figure~\ref{figure:ee-ZHH-frames}~\footnote{Figures~\ref{figure:ee-ZHH-frames}, \ref{figure:ee-ZHH-diagrams}, \ref{figure:ee-ZHH-CP}, \ref{figure:ee-ZHH-CPT} are drawn by using the program JaxoDraw~\cite{Binosi:2003yf}.}. The four-momentum and helicity of each particle are shown in parenthesis. A single object as the sum of the two Higgs bosons is represented by $2H$, whose four-momentum is $q^{\mu}_{}$. We choose the direction of $\vec{q}$ as the $z$-axis and the $\vec{p}_1^{} \times \vec{k}$ direction as the $y$-axis. 
The scattering $p_1^{}+p_2^{} \to k + q$ takes place in the $x$-$z$ plane. 
The polar angle of the $Z$ boson from the electron momentum direction is denoted by $\Theta$. Because we neglect the $e^{\pm}_{}$ masses {\it and} $e^-_{}$ and $e^+_{}$ always construct a four-vector in our process, the helicity of $e^+_{}$ is always opposite to that of $e^-_{}$. 
In this coordinate system, the four-momenta can be parametrized as follows:
\begin{align}
p^{\mu}_{} & \equiv (p^{}_{1} + p^{}_{2})^{\mu}_{} = (E, 0, 0, 0 ), \nonumber \\
p^{\mu}_{1} & = \frac{E}{2}\bigl( 1, \sin{\Theta}, 0, -\cos{\Theta} \bigr),\nonumber \\
p^{\mu}_{2} & = \frac{E}{2}\bigl( 1, -\sin{\Theta}, 0, \cos{\Theta} \bigr),\nonumber \\
k^{\mu}_{} & = ( w, 0, 0, -l ),\nonumber \\
q^{\mu}_{} & = (q^{}_1+q^{}_2)^{\mu}_{} = ( E-w, 0, 0, l),
\end{align}
where $E$ is the $e^+_{}e^-_{}$ c.m. energy, 
$w$ is the energy of the $Z$ boson: $w=(E^2_{} + m_Z^2 - Q^2_{})/(2E)$ where $Q^2_{}=q \cdot q$, $-l$ is the momentum of the $Z$ boson: $l=\sqrt{w^2_{}-m_Z^2}$, and $q^{\mu}_{1,2}$ are the four-momenta of the two Higgs bosons. 
We parametrize $q^{\mu}_{1,2}$ in the rest frame of $q^{\mu}_{}$ as
\begin{align}
q^{\mu}_{} &= ( q^{}_{1} + q^{}_{2} )^{\mu}_{} = ( Q, 0, 0, 0), \nonumber \\
q^{\mu}_{1} &= \bigl( Q/2, r \sin{\xi} \cos{\phi}, r \sin{\xi} \sin{\phi}, r \cos{\xi} \bigr), \nonumber \\
q^{\mu}_{2} &= \bigl( Q/2, - r \sin{\xi} \cos{\phi}, - r \sin{\xi} \sin{\phi}, - r \cos{\xi} \bigr), \label{eq:momentumQq1q2}
\end{align}
where $r=\sqrt{Q^2_{}/4-m_H^2}$. Since we cannot distinguish the two Higgs bosons, we define the regions of the angles as $0 \le \xi \le \pi/2$ and $0 \le \phi \le 2\pi$ and identify the Higgs boson whose momentum along the $z$-axis is positive as the Higgs boson that has $q_1^{\mu}$. 
The four-momenta $q^{\mu}_{1,2}$ in our $e^+_{}e^-_{}$ c.m. frame can be easily obtained by a single boost along the positive direction of the $z$-axis:
\begin{subequations}\label{eq:q1andq2sum}
\begin{align}
q^{\mu}_{1} &= \biggl( \frac{E-w}{2}+ \frac{l}{Q}r \cos{\xi}, r \sin{\xi} \cos{\phi}, r \sin{\xi} \sin{\phi}, \frac{l}{2} + \frac{E-w}{Q} r \cos{\xi} \biggr), \label{eq:q1} \\
q^{\mu}_{2} &= \biggl( \frac{E-w}{2} - \frac{l}{Q}r \cos{\xi}, - r \sin{\xi} \cos{\phi}, - r \sin{\xi} \sin{\phi}, \frac{l}{2} - \frac{E-w}{Q} r \cos{\xi} \biggr). \label{eq:q2}
\end{align}
\end{subequations}
We introduce the four-momenta $t^{\mu}_{}$ and $u^{\mu}_{}$ of the intermediate $Z$ boson or the photon $\gamma$ in the diagrams (2) and (3) of Figure~\ref{figure:ee-ZHH-diagrams}, respectively:
\begin{align} \label{eq:tandu}
t^{\mu}_{} & = ( k + q_1^{} )^{\mu}_{} \nonumber \\
& = \biggl( \frac{E+w}{2}+ \frac{l}{Q}r \cos{\xi}, r \sin{\xi} \cos{\phi}, r \sin{\xi} \sin{\phi}, -\frac{l}{2} + \frac{E-w}{Q} r \cos{\xi} \biggr), \nonumber \\
u^{\mu}_{} & = ( k + q_2^{} )^{\mu}_{} \nonumber \\
& = \biggl( \frac{E+w}{2} - \frac{l}{Q}r \cos{\xi}, -r \sin{\xi} \cos{\phi}, - r \sin{\xi} \sin{\phi}, -\frac{l}{2} - \frac{E-w}{Q} r \cos{\xi} \biggr). 
\end{align}

\section{Differential cross section}\label{observables}

\begin{figure}[t]
\centering
\includegraphics[scale=0.45]{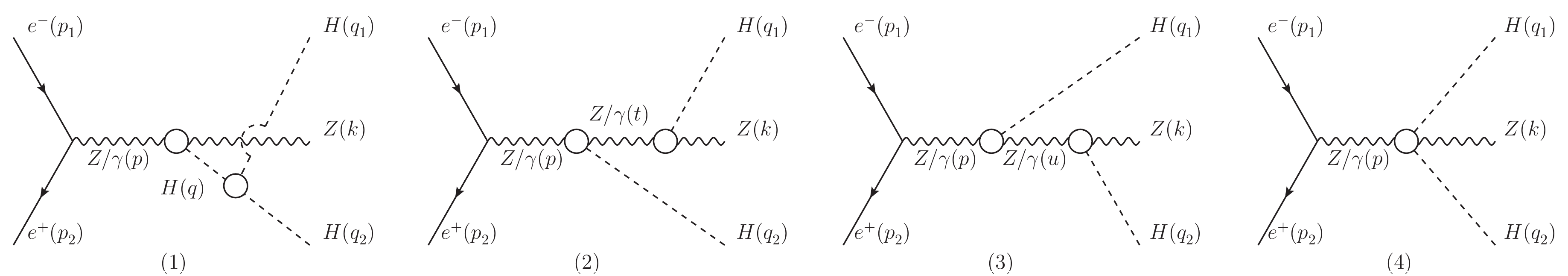}
\caption{\small Feynman diagrams for $e^+_{} e^-_{} \to ZHH$ with our effective Lagrangian eq~(\ref{eq:lagrangian}). The four-momentum of each particle is shown in parenthesis.}
\label{figure:ee-ZHH-diagrams}
\end{figure}

We present an analytic expression for the differential cross section by assuming non-standard Higgs couplings. The effective Lagrangian, from which we obtain the Higgs couplings, will be given in Section~\ref{numerical}. 
The Feynman diagrams contributing to the scattering amplitudes of the process $e^+_{}e^-_{} \to ZHH$ with our non-standard Higgs couplings are shown in Figure~\ref{figure:ee-ZHH-diagrams}. It is an easy task to derive the scattering amplitudes in terms of the kinematic variables defined in Section~\ref{kinematics}. 
We find that the amplitude-squared summed over the $Z$ boson helicity $\lambda$ for a given electron helicity $\tau$ has the following form in the most general case:
\begin{align}
\sum_{\lambda=\pm,0} | {\cal M}^{\lambda}_{\tau} |^2_{} 
= & 
F_{1}^{} \bigl(1+\cos^2_{}{\Theta} \bigr)
+ F_{2}^{} \bigl(1-3\cos^2_{}{\Theta} \bigr)
+ F_{3}^{}  \cos{\Theta} 
 + F_{4}^{}  \sin{\Theta} \cos{\phi} 
 + F_{5}^{} \sin{2\Theta} \cos{\phi}  \nonumber  \\
& + F_{6}^{} \sin^2_{}{\Theta} \cos{2\phi}
 + F_{7}^{}  \sin{\Theta} \sin{\phi} 
+ F_{8}^{} \sin{2\Theta} \sin{\phi} 
+ F_{9}^{} \sin^2_{}{\Theta} \sin{2\phi} 
, \label{eq:amplitude-squared}
\end{align}
where the $\Theta$ and $\phi$ dependences are completely factorized and the 9 functions $F_i^{}$ as the angular coefficients are independent on these two angles: $F_i^{} = F_i^{}(\tau, Q, \xi)$~\footnote{The functions $F_i^{}$ depend on the $e^+_{}e^-_{}$ c.m. energy (i.e. $E$ in our notation), hence $F_i^{} = F_i^{}(\tau, Q, \xi, E)$ is the more appropriate expression. However, we regard $E$ as a fixed value and do not write $E$ explicitly in the arguments of functions throughout the paper.}. The 9 functions $F_i^{}$ depend on the Higgs couplings. The 9 functions $F_i^{}$ can be experimentally determined by measuring the angles $\Theta$ and $\phi$, therefore used to study the Higgs couplings. For an approach to isolate the functions, see eqs.~(3.28) and (3.29) of ref.~\cite{Nakamura:2017ihk}. 
By means of eq.~(\ref{eq:amplitude-squared}), the complete differential cross section for a given electron helicity $\tau$ is given by
\begin{align}
\frac{d \sigma(\tau)}{ d \Omega}
\equiv
\frac{d \sigma(\tau)}{ d \cos{\Theta} dQ^2_{} d \cos{\xi} d \phi}
=
\frac{1}{1024 \pi^4_{}} \frac{l}{E^3_{}} \sqrt{1-\frac{4m_H^2}{Q^2_{}}} \sum_{\lambda=\pm,0} | {\cal M}^{\lambda}_{\tau} |^2_{}.\label{eq:comp-diff}
\end{align}
By performing the integration over $\phi$, we obtain the analytic form of the $\cos{\Theta}$ distribution:
\begin{align}
\int^{2\pi}_{0} d\phi \frac{d \sigma(\tau)}{ d \Omega}
=
\frac{1}{512 \pi^3_{}} \frac{l}{E^3_{}} \sqrt{1-\frac{4m_H^2}{Q^2_{}}}
\Bigl[ F_{1}^{} \bigl(1+\cos^2_{}{\Theta} \bigr)
+ F_{2}^{} \bigl(1-3\cos^2_{}{\Theta} \bigr)
+ F_{3}^{} \cos{\Theta} \Bigr],\label{eq:cosdist1}
\end{align}
where the other 6 terms were eliminated by the integration. The numerical studies of the $\cos{\Theta}$ distribution in the literature e.g.~\cite{Miller:1999ct, Kumar:2014zra} actually probe the 3 coefficients, which can be obtained by integrating eq.~(\ref{eq:cosdist1}) over $Q^2_{}$ and $\cos{\xi}$: 
\begin{align}
 \frac{d \sigma(\tau)}{d \cos{\Theta}} 
& = \int^{(E-m_Z^{})^2_{}}_{4m_H^2} dQ^2_{} \int^1_0 d \cos{\xi} 
\int^{2\pi}_{0} d\phi \frac{d \sigma(\tau)}{ d \Omega} \nonumber \\
& =
\int^{(E-m_Z^{})^2_{}}_{4m_H^2} dQ^2_{} \int^1_0 d \cos{\xi} 
\frac{1}{512 \pi^3_{}} \frac{l}{E^3_{}} \sqrt{1-\frac{4m_H^2}{Q^2_{}}}
\Bigl[ F_{1}^{} \bigl(1+\cos^2_{}{\Theta} \bigr)
+ F_{2}^{} \bigl(1-3\cos^2_{}{\Theta} \bigr)
+ F_{3}^{} \cos{\Theta} \Bigr]. \label{eq:cosdist2}
\end{align}
By further integrating eq.~(\ref{eq:cosdist1}) over $\cos{\Theta}$, we obtain
\begin{align}
\frac{d \sigma(\tau)}{ dQ^2_{} d \cos{\xi}}
=
\int^1_{-1} d \cos{\Theta} \int^{2\pi}_{0} d\phi \frac{d \sigma(\tau)}{ d \Omega}
= \frac{1}{192 \pi^3_{}} \frac{l}{E^3_{}} \sqrt{1-\frac{4m_H^2}{Q^2_{}}}
F_{1}^{},\label{eq:dQ2dcosxi}
\end{align}
where the other 2 terms were eliminated by the integration. From this, we can easily obtain the analytic form of the $Q^2_{}$ distribution which has been numerically studied e.g. in refs.~\cite{Battaglia:2001nn, Baur:2009uw} and that of the $\cos{\xi}$ distribution which ref.~\cite{Boudjema:1995cb} mentions can be a good observable for measuring $\lambda_H^{}$:
\begin{subequations}\label{eq:dQ2anddcosxi}
\begin{align}
\frac{d \sigma(\tau)}{ dQ^2_{} }
&= \int^1_0 d \cos{\xi} \frac{1}{192 \pi^3_{}} \frac{l}{E^3_{}} \sqrt{1-\frac{4m_H^2}{Q^2_{}}}
F_{1}^{}, \\
\frac{d \sigma(\tau)}{d \cos{\xi}}
&= \int^{(E-m_Z^{})^2_{}}_{4m_H^2} dQ^2_{} \frac{1}{192 \pi^3_{}} \frac{l}{E^3_{}} \sqrt{1-\frac{4m_H^2}{Q^2_{}}}
F_{1}^{}.
\end{align}
\end{subequations}
The total cross section for a given electron helicity $\tau$ is given by
\begin{align}
\sigma(\tau) = \int^{(E-m_Z^{})^2_{}}_{4m_H^2} dQ^2_{} \int^1_0 d \cos{\xi} \frac{1}{192 \pi^3_{}} \frac{l}{E^3_{}} \sqrt{1-\frac{4m_H^2}{Q^2_{}}}
F_{1}^{}.\label{eq:totalcrosssection}
\end{align}
We introduce the scaling variables
\begin{align}
x_1^{} = \frac{2 q_1^0}{E}, \ \
x_2^{} = \frac{2 q_2^0}{E}.
\end{align}
Note that $q_{1}^0$ and $q_{2}^0$ are defined in eq.~(\ref{eq:q1andq2sum}). 
By straightforward variable conversions in eq.~(\ref{eq:dQ2dcosxi}), we obtain
\begin{align}
\frac{d \sigma(\tau)}{ dx_1^{} dx_2^{}}
= \frac{1}{192 \pi^3_{}}
F_{1}^{} \label{eq:dx1dx2}
\end{align}
which has been derived in refs.~\cite{Djouadi:1996ah, Osland:1998hv, Djouadi:1999gv} and
\begin{align}
\frac{d \sigma(\tau)}{ dQ_{}^2 d(t\cdot t)}
= \frac{1}{192 \pi^3_{} E^4_{}}
F_{1}^{} \label{eq:dQ2dt2}
\end{align}
which has been derived in ref.~\cite{Contino:2013gna}. Note that the phase space region $0 \le \xi \le \pi/2$ (see below eq.~(\ref{eq:momentumQq1q2}) corresponds to $x_1^{} \ge x_2^{}$ and $t \cdot t \ge u \cdot u$. The four-momenta $t^{\mu}_{}$ and $u^{\mu}_{}$ are defined in eq.~(\ref{eq:tandu}). We emphasize that the observables which exist in the literature and are re-derived above in eqs.~(\ref{eq:dQ2anddcosxi}), (\ref{eq:totalcrosssection}), (\ref{eq:dx1dx2}) and (\ref{eq:dQ2dt2}) are directly related to the function $F_1^{}$, which is just one of the 9 functions in the differential cross section.

\section{Symmetry properties}\label{symmetry}

\begin{figure}[t]
\centering
\includegraphics[scale=0.5]{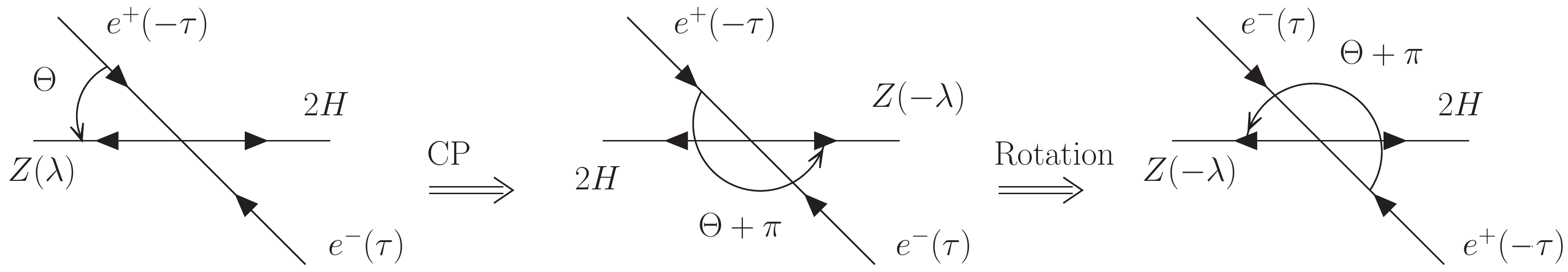}
\caption{\small The states after the CP transformation are shown. At the second step, a rotation around the $y$-axis by $\pi$ is performed. The helicity of each particle is shown in parenthesis.}
\label{figure:ee-ZHH-CP}
\end{figure}

The conditions imposed by symmetries lead to constraints on some of the 9 functions $F_i^{}$. 
The picture in Figure~\ref{figure:ee-ZHH-CP} shows the original states and the states after the charge-conjugation (C) and parity (P) transformation. 
After the CP transformation, the states are simply rotated around the $y$-axis by $\pi$ 
and we make the four-momentum $q^{\mu}_{}$ come back to the original position. Note that we always have a freedom of performing 3-dimensional spatial rotations. 
While $q^{\mu}_{}$ is unchanged, the four-momenta $q_1^{\mu}$ and $q_2^{\mu}$ are changed by the CP transformation and the rotation as
\begin{align}
q_1^{\mu} 
& = \biggl( \frac{E-w}{2}+ \frac{l}{Q}r \cos{\xi},  r \sin{\xi} \cos{\phi},  r \sin{\xi} \sin{\phi}, \frac{l}{2} + \frac{E-w}{Q} r \cos{\xi} \biggr), \nonumber \\
&  \xrightarrow[\mathrm{CP}]{} 
\biggl( \frac{E-w}{2}+ \frac{l}{Q}r \cos{\xi}, - r \sin{\xi} \cos{\phi}, - r \sin{\xi} \sin{\phi}, - \frac{l}{2} - \frac{E-w}{Q} r \cos{\xi} \biggr), \nonumber \\
& \xrightarrow[\mathrm{Rotation}]{}  \biggl( \frac{E-w}{2}+ \frac{l}{Q}r \cos{\xi},  r \sin{\xi} \cos{\phi}, - r \sin{\xi} \sin{\phi},  \frac{l}{2} + \frac{E-w}{Q} r \cos{\xi} \biggr).\label{eq:q1afterTrans}
\end{align}
Notice that only the $y$-component of $q_1^{\mu}$ changes sign, which indicates that the azimuthal angle $\phi$ is $2\pi-\phi$ after the transformations. Therefore, CP invariance leads to the following relation for the differential cross section:
\begin{align}
d\sigma(\tau, \Theta, Q, \xi, \phi) = d\sigma(\tau, \Theta+\pi, Q, \xi, 2\pi-\phi),\label{eq:CPrelation}
\end{align}
where the one on the left hand side corresponds to the original states shown in the the left picture of Figure~\ref{figure:ee-ZHH-CP} and the one on the right hand side corresponds to the states after the CP transformation and the rotation shown in the the right picture of Figure~\ref{figure:ee-ZHH-CP}. 
The explicit form of the right hand side of this equation is, from eqs.~(\ref{eq:amplitude-squared}) and (\ref{eq:comp-diff}), given by
\begin{align}
\frac{d \sigma(\tau)}{ d \cos{\Theta} dQ^2_{} d \cos{\xi} d \phi}
= & \frac{1}{1024 \pi^4_{}} \frac{l}{E^3_{}} \sqrt{1-\frac{4m_H^2}{Q^2_{}}} \Bigl[
 F_{1}^{} \bigl(1+\cos^2_{}{\Theta} \bigr)
+ F_{2}^{} \bigl(1-3\cos^2_{}{\Theta} \bigr)
- F_{3}^{}  \cos{\Theta} \nonumber \\
& - F_{4}^{} \sin{\Theta} \cos{\phi} 
 + F_{5}^{} \sin{2\Theta} \cos{\phi}  
 + F_{6}^{} \sin^2_{}{\Theta} \cos{2\phi} \nonumber \\
& + F_{7}^{} \sin{\Theta} \sin{\phi} 
- F_{8}^{} \sin{2\Theta} \sin{\phi} 
- F_{9}^{} \sin^2_{}{\Theta} \sin{2\phi} \Bigr]
.\label{eq:crossection-CP}
\end{align}
Let us remind that $F_i^{}$ are independent on $\Theta$ and $\phi$, thus they have the same forms in the both sides of eq.~(\ref{eq:CPrelation})~\footnote{We actually perform the rotation in order to make all of $F_i^{}$ invariant. Some of $F_i^{}$ are not invariant without the rotation. Even without the rotation, however, our conclusion that the $F_3^{}$, $F_4^{}$, $F_8^{}$ and $F_9^{}$ terms are CP-odd should remain the same, as long as physics is invariant under 3-dimensional spatial rotations.}. In eq.~(\ref{eq:crossection-CP}), we observe that the 4 terms of the 9 terms change sign. These terms, namely the $F_3^{}$, $F_4^{}$, $F_8^{}$ and $F_9^{}$ terms, are CP-odd. The 4 functions $F_3^{}$, $F_4^{}$, $F_8^{}$ and $F_9^{}$ must be zero if CP is conserved. In other words, observation of non-zero values in these 4 functions signals CP non-conservation. \\

\begin{figure}[t]
\centering
\includegraphics[scale=0.5]{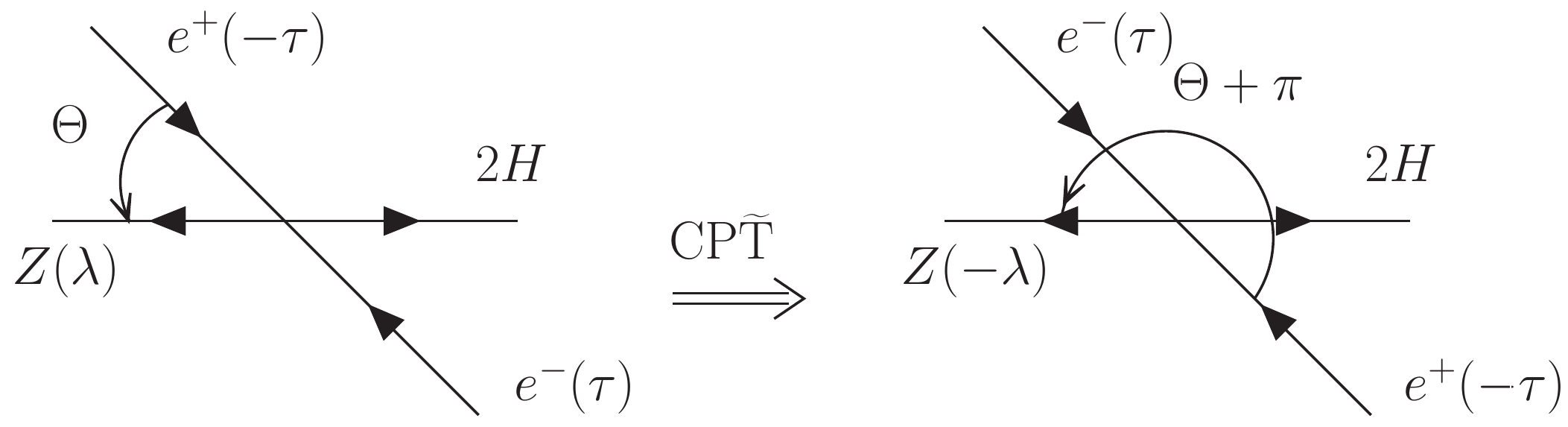}
\caption{\small The states after the CP and time-reversal transformation without interchanging the initial and final states ($\mathrm{\widetilde{T}}$) are shown. The helicity of each particle is shown in parenthesis.}
\label{figure:ee-ZHH-CPT}
\end{figure}

Secondly, the picture in Figure~\ref{figure:ee-ZHH-CPT} shows the original states and the states after the C, P and time-reversal transformation without interchanging the initial and final states (i.e. it does not reverse the time flow from the initial state to the final state). We denote it by $\mathrm{\widetilde{T}}$. $\mathrm{CP\widetilde{T}}$ invariance leads to the following relation for the differential cross section:
\begin{align}
d\sigma(\tau, \Theta, Q, \xi, \phi) = d\sigma(\tau, \Theta+\pi, Q, \xi, \phi),\label{eq:CPTrelation}
\end{align}
where the one on the left hand side corresponds to the original states shown in the the left picture of Figure~\ref{figure:ee-ZHH-CPT} and the one on the right hand side corresponds to the states after the $\mathrm{CP\widetilde{T}}$ transformation shown in the right picture of Figure~\ref{figure:ee-ZHH-CPT}. 
The explicit form of the right hand side of this equation is, from eqs.~(\ref{eq:amplitude-squared}) and (\ref{eq:comp-diff}), given by
\begin{align}
\frac{d \sigma(\tau)}{ d \cos{\Theta} dQ^2_{} d \cos{\xi} d \phi}
= & \frac{1}{1024 \pi^4_{}} \frac{l}{E^3_{}} \sqrt{1-\frac{4m_H^2}{Q^2_{}}} \Bigl[
 F_{1}^{} \bigl(1+\cos^2_{}{\Theta} \bigr)
+ F_{2}^{} \bigl(1-3\cos^2_{}{\Theta} \bigr)
- F_{3}^{}  \cos{\Theta} \nonumber \\
& - F_{4}^{} \sin{\Theta} \cos{\phi} 
 + F_{5}^{} \sin{2\Theta} \cos{\phi}  
 + F_{6}^{} \sin^2_{}{\Theta} \cos{2\phi} \nonumber \\
& - F_{7}^{} \sin{\Theta} \sin{\phi} 
+ F_{8}^{} \sin{2\Theta} \sin{\phi} 
+ F_{9}^{} \sin^2_{}{\Theta} \sin{2\phi} \Bigr], \label{eq:crossection-CPT}
\end{align}
where we observe that the 3 terms of the 9 terms change sign. These terms, namely the $F_3^{}$, $F_4^{}$, and $F_7^{}$ terms, are $\mathrm{CP\widetilde{T}}$-odd. The 4 functions $F_3^{}$, $F_4^{}$, and $F_7^{}$ must be zero if $\mathrm{CP\widetilde{T}}$ is conserved. In other words, observation of non-zero values in these 3 functions signals $\mathrm{CP\widetilde{T}}$ violation, which indicates the existence of re-scattering effects~\cite{Hagiwara:1986vm}. In Table~\ref{table:symproperty}, we summarize the symmetry properties of the functions. Notice that $F_3^{}$ and $F_4^{}$ are both CP-odd and $\mathrm{CP\widetilde{T}}$-odd. Once we experimentally confirm that both the CP-odd functions (i.e. $F_8^{}$ and $F_9^{}$) and the $\mathrm{CP\widetilde{T}}$-odd function (i.e. $F_7^{}$) are small, we may ignore $F_3^{}$ and $F_4^{}$ since these are doubly suppressed.

\begin{table}
\begin{tabular}{@{}c cccccc@{}} 
\toprule


Functions  & \multicolumn{2}{l}{Symm. properties}  & beam pol.\\  

\cmidrule(l){2-3}

 &  CP & $\mathrm{CP\widetilde{T}}$ &  \\ 

\midrule


$F_1^{}$ & $+$ & $+$  & - \\

$F_2^{}$ & $+$ & $+$  & - \\

$F_3^{}$ & $-$ & $-$  & $\circ$ \\

$F_4^{}$ & $-$ & $-$  & $\circ$ \\

$F_5^{}$ & $+$ & $+$  & - \\

$F_6^{}$ & $+$ & $+$  & - \\

$F_7^{}$ & $+$ & $-$  & $\circ$ \\

$F_8^{}$ & $-$ & $+$  & - \\

$F_9^{}$ & $-$ & $+$  & - \\


\bottomrule

\end{tabular}
\caption{\small Symmetry properties of the 9 functions in the differential cross section. 
The symbol $+$ ($-$) means that the function is even (odd) under CP or $\mathrm{CP\widetilde{T}}$. Observation of non-zero values in the CP-odd functions signals CP non-conservation. Observation of non-zero values in the $\mathrm{CP\widetilde{T}}$-odd functions indicates the existence of re-scattering effects. 
The symbol $\circ$ in the last column indicates that the function can be suppressed without polarized $e^+_{}e^-_{}$ beams.}
\label{table:symproperty}
\end{table}

\section{Numerical studies}\label{numerical}

We obtain non-standard Higgs couplings to the Higgs boson itself, the $Z$ boson and the photon $\gamma$ from the following effective Lagrangian~\cite{Hagiwara:1993sw}:
\begin{align}
& {\cal L}_{\mathrm{eff}}^{} 
 =  \bigl( 1 + \delta_1^{} ) m_Z^{2} \frac{H}{v} Z_{\mu}^{}Z^{\mu}_{}
+ 
\sum_{V=Z,A} \Bigl\{
\delta_2^{V}  \frac{H}{v} Z_{\mu\nu}^{} V^{\mu\nu}_{} +
\delta_3^{V} \frac{1}{v} \bigl[ (\partial^{\mu}_{}H) Z^{\nu}_{} - (\partial^{\nu}_{}H) Z^{\mu}_{} \bigr] V_{\mu\nu}^{}  
+ \tilde{\delta}_4^{V} \frac{H}{v} Z_{\mu\nu}^{} \widetilde{V}^{\mu\nu}_{}
\Bigr\} 
\nonumber \\
&+ \bigl( 1 + \delta_5^{} ) m_Z^{2} \frac{H^2_{}}{2v^2_{}} Z_{\mu}^{}Z^{\mu}_{}
+ 
 \sum_{V=Z,A} \Bigl\{
 \delta_6^{V} \frac{H^2_{}}{2v^2_{}} Z_{\mu\nu}^{} V^{\mu\nu}_{} +
\delta_7^{V} \frac{H}{v^2_{}} \bigl[ (\partial^{\mu}_{}H) Z^{\nu}_{} - (\partial^{\nu}_{}H) Z^{\mu}_{} \bigr] V_{\mu\nu}^{}  
+ \tilde{\delta}_8^{V} \frac{H^2_{}}{2v^2_{}}  Z_{\mu\nu}^{} \widetilde{V}^{\mu\nu}_{}
\Bigr\} 
\nonumber \\ 
& +
\delta_2^{AA} \frac{H}{v} A_{\mu\nu}^{} A^{\mu\nu}_{} 
+ \tilde{\delta}_4^{AA} \frac{H}{v} A_{\mu\nu}^{} \widetilde{A}^{\mu\nu}_{}
 - \frac{m_H^2}{2v}( 1 + \delta_9^{} )H^3_{} + 
\delta_{10}^{} \frac{H}{v} (\partial^{\mu}_{}H)^2_{},
\label{eq:lagrangian}
\end{align}
where $V_{\mu\nu}^{} = \partial_{\mu}^{} V_{\nu}^{} - \partial_{\nu}^{} V_{\mu}^{}$, $\widetilde{V}^{\mu\nu}_{}=\frac{1}{2} \epsilon^{\mu\nu\rho\sigma}_{} V_{\rho\sigma}^{}$ with our convention $\epsilon_{0123}^{}=+1$, and $v$ is the vacuum expectation value of the Higgs field: $v^{-2}_{}=\sqrt{2}G_F^{}$. 
All of the 18 coefficients $\delta_i^{}$ are zero at the tree level in the SM. The 5 operators whose coefficients are $\tilde{\delta}_4^{V}$, $\tilde{\delta}_8^{V}$ and $\tilde{\delta}_4^{AA}$ are CP-odd and the other 13 operators are CP-even. If both the CP-even operator(s) and the CP-odd operator(s) exist, the theory is not CP conserving. If we consider the $SU(2)\times U(1)$ gauge invariant dimension six operators which consists of the gauge boson fields and the Higgs doublet field, there are 8 CP-even operators and 5 CP-odd operators~\cite{Hagiwara:1993sw, Hagiwara:1993ck} which contribute to the Higgs couplings relevant to our process and their effects can be expressed as their contributions to our 18 coefficients $\delta_i^{}$. In addition, 2 CP-even operators affect our process through the renormalization of the SM parameters and the external $Z$ and Higgs fields~\cite{Hagiwara:1993sw}. Feynman diagrams for the process $e^+_{}e^-_{}\to ZHH$ with this effective Lagrangian have been already shown in Figure~\ref{figure:ee-ZHH-diagrams}. We have 12 diagrams maximum.\\

We integrate the differential cross section in eq.~(\ref{eq:comp-diff}) over $Q^2_{}$ and $\cos{\xi}$:
\begin{align}
\frac{d \sigma(\tau)}{ d \cos{\Theta} d \phi}
&= {\cal F}_{1}^{} \bigl(1+\cos^2_{}{\Theta} \bigr)
+ {\cal F}_{2}^{} \bigl(1-3\cos^2_{}{\Theta} \bigr)
+ {\cal F}_{3}^{}  \cos{\Theta} 
 + {\cal F}_{4}^{} \sin{\Theta} \cos{\phi}  
+ {\cal F}_{5}^{} \sin{2\Theta} \cos{\phi} \nonumber \\
&+ {\cal F}_{6}^{} \sin^2_{}{\Theta} \cos{2\phi} 
 + {\cal F}_{7}^{} \sin{\Theta} \sin{\phi} 
+ {\cal F}_{8}^{} \sin{2\Theta} \sin{\phi} 
+ {\cal F}_{9}^{} \sin^2_{}{\Theta} \sin{2\phi},\label{eq:angular-dist}
\end{align}
where
\begin{align}
{\cal F}_i^{}(\tau) =  \int^{(E-m_Z^{})^2_{}}_{4m_H^2} dQ^2_{} \int^1_0 d \cos{\xi} \frac{1}{1024\pi^4_{}} \frac{l}{E^3_{}} \sqrt{1-\frac{4m_H^2}{Q^2_{}}} F_i^{} (\tau, Q, \xi). \label{eq:IntegratedF}
\end{align}
Let us remind that the total cross section is directly related to ${\cal F}_1^{}(\tau)$:
\begin{align}
\sigma(\tau) = \frac{16}{3}\pi {\cal F}_1^{}(\tau).\label{eq:F1andCross}
\end{align}
We assume unpolarized $e^+_{}e^-_{}$ beams and define observables as 
\begin{align}
A_i^{} \equiv \frac{\sum_{\tau}{\cal F}_i^{}(\tau)}{\sum_{\tau}{\cal F}_1^{}(\tau)}.\label{eq:ObservableA}
\end{align}
The symmetry property of $A_i^{}$ is the same as that of the corresponding function $F_i^{}$. We note the advantages of the observables $A_i^{}$:
\begin{itemize}
\item Some of systematic uncertainties such as the luminosity uncertainty cancel.

\item We expect that ${\cal F}_i^{}$ $(i=2,3,\cdots,9)$ depend on the Higgs couplings in different ways from ${\cal F}_1^{}$ (i.e. the total cross section) so that ${\cal F}_i^{}$ $(i=2,3,\cdots,9)$ provide us different information on the Higgs couplings. The observables $A_i^{}$ have the form that is sensitive to the difference between ${\cal F}_i^{}$ $(i=2,3,\cdots,9)$ and ${\cal F}_1^{}$ in the dependence on the Higgs couplings.

\end{itemize}
For our numerical results, we set $E=500$ GeV, $m_Z^{}=91.188$ GeV, $\Gamma_Z^{}=2.5$ GeV, $m_H^{}=125.5$ GeV, $\Gamma_H^{}=0$ GeV and $e=\sqrt{4\pi\alpha}$ with $\alpha=1/128$. We assume that the $Z$ boson and the Higgs bosons can be reconstructed. The phase space integration is performed with the program BASES~\cite{Kawabata:1995th}. \\

\begin{figure}[t]
\centering
\includegraphics[scale=0.52]{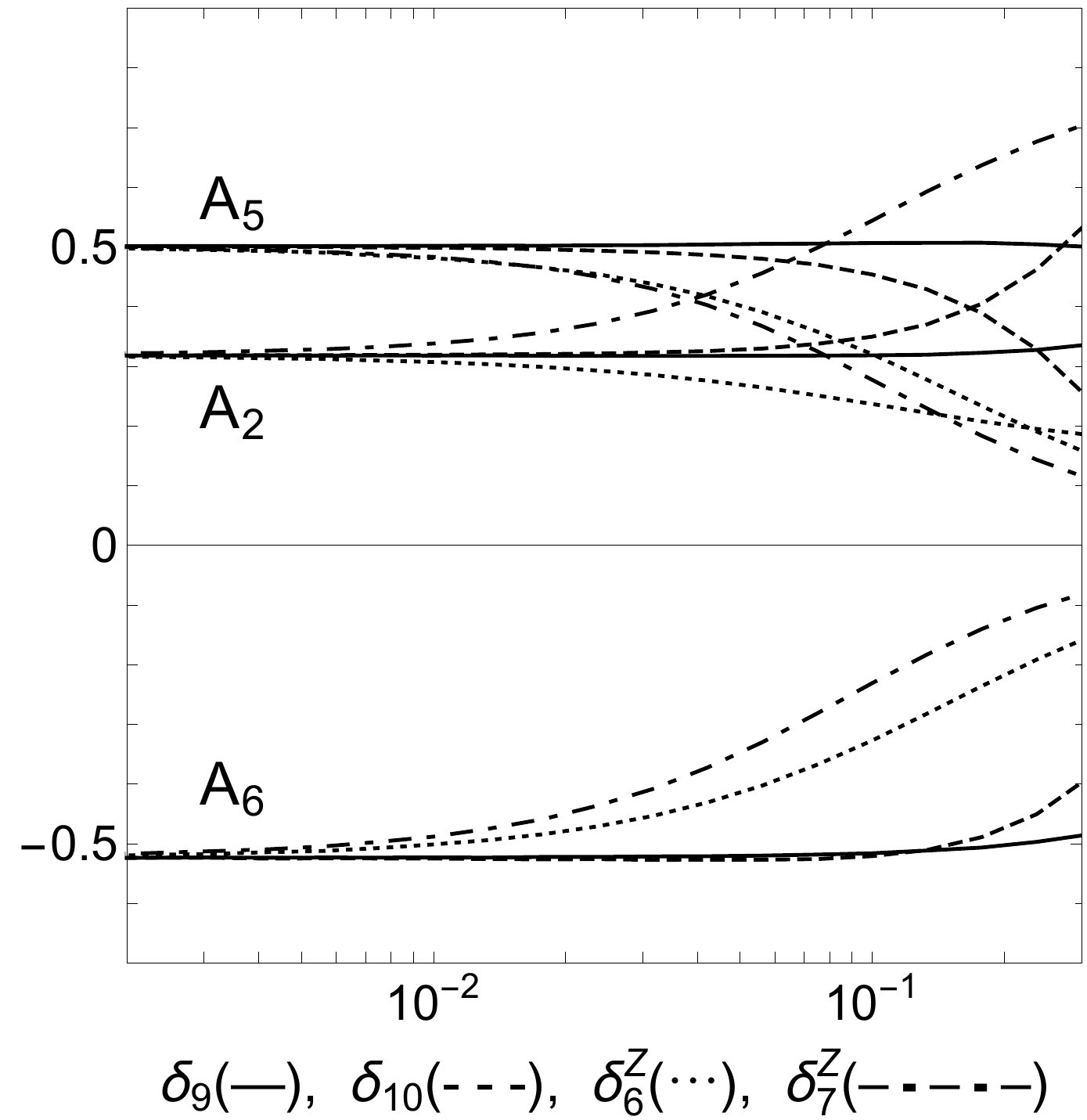}
\includegraphics[scale=0.52]{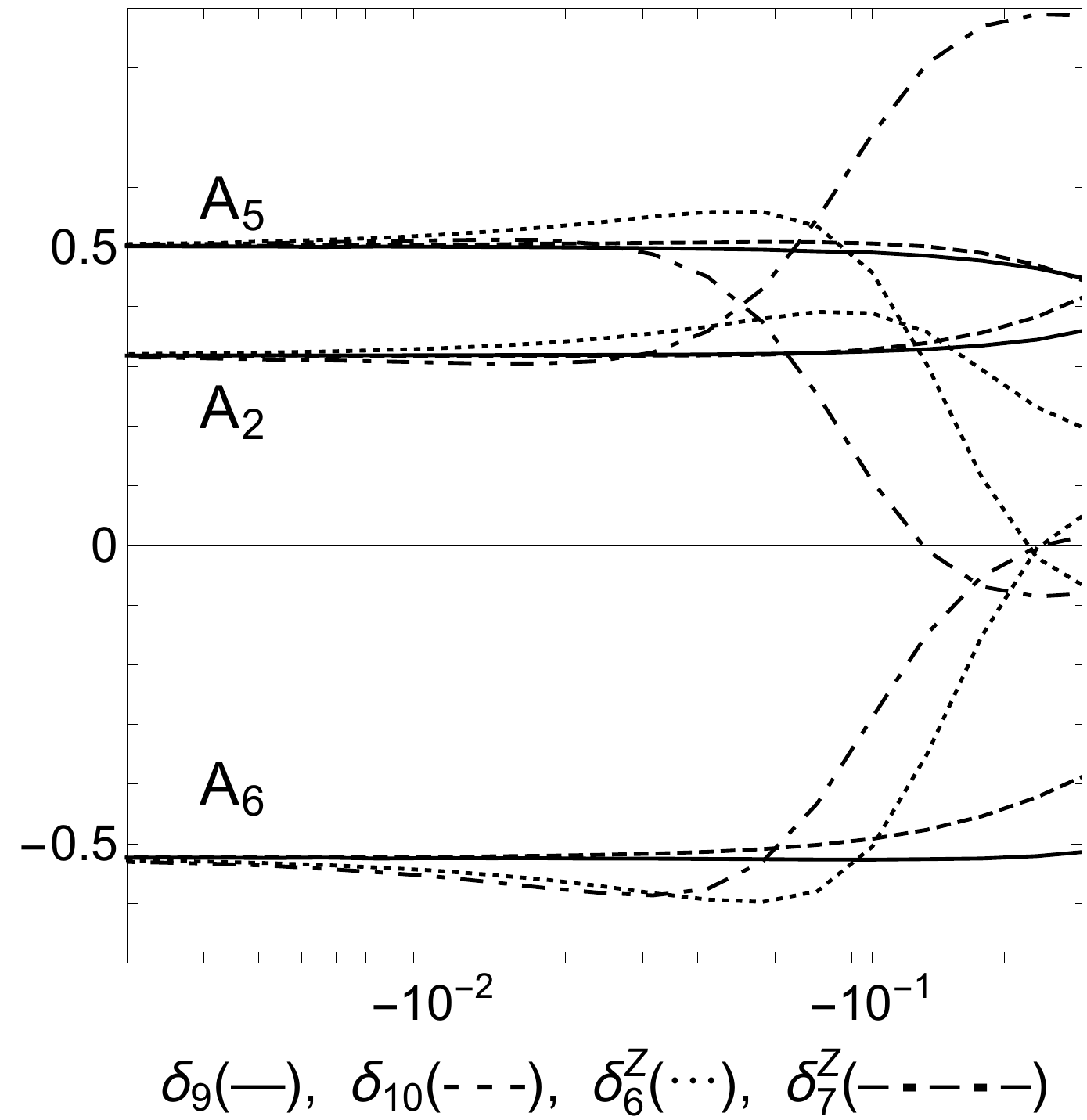}
\caption{\small $A_2^{}$, $A_5^{}$ and $A_6^{}$ are shown as deviations caused by adding non-zero parameters $\delta_9^{}$ (solid curve), $\delta_{10}^{}$ (dashed curve), $\delta_6^Z$ (dotted curve) and $\delta_7^{Z}$ (broken curve). In the left panel the parameters take positive values, and in the right panel the parameters take negative values.}
\label{figure:ee-ZHH-fig1}
\end{figure}

\begin{figure}[t]
\centering
\includegraphics[scale=0.52]{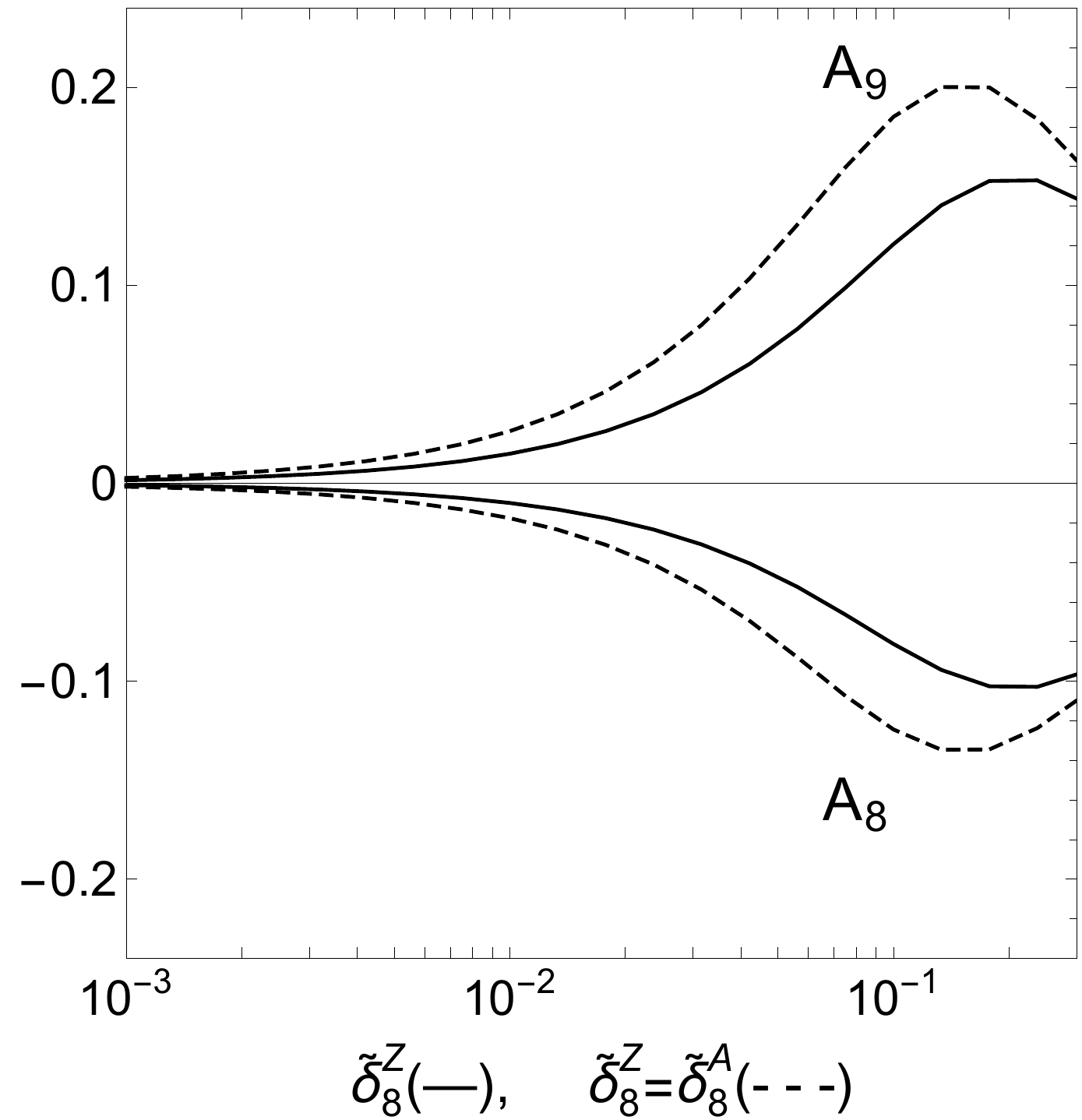}
\includegraphics[scale=0.52]{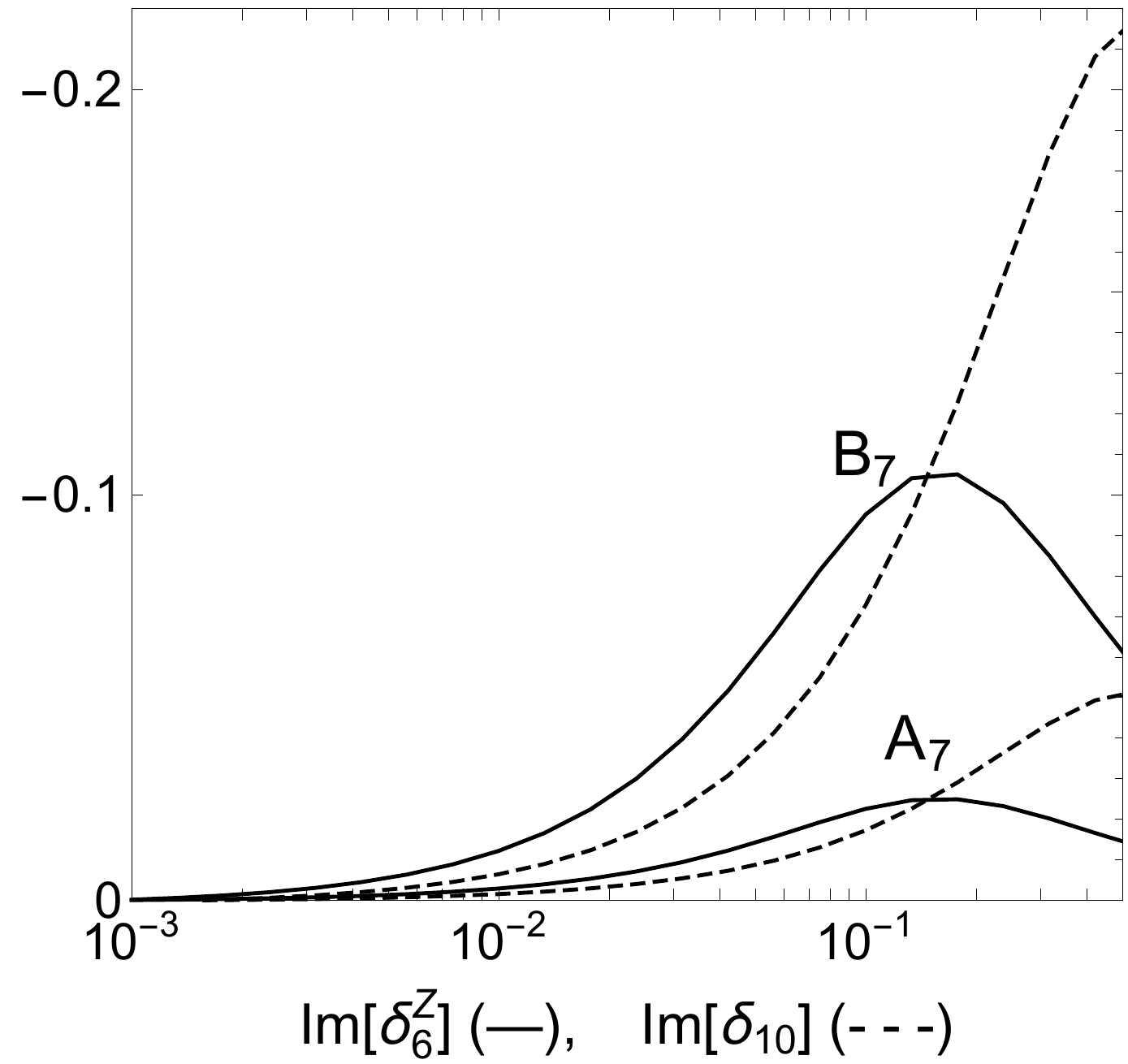}
\caption{\small {\it Left:} CP-odd observables $A_8^{}$ and $A_9^{}$ are shown as deviations caused by adding non-zero parameters $\tilde{\delta}_8^{Z}$ (solid curve) and $\tilde{\delta}_8^{Z}=\tilde{\delta}_8^{A}$ (dashed curve). {\it Right:} $\mathrm{CP\widetilde{T}}$-odd observables $A_7^{}$ and $B_7^{}$ are shown as deviations caused by adding non-zero imaginary parts in the parameters $\delta_6^Z$ (solid curve) and $\delta_{10}^{}$ (dashed curve).}
\label{figure:ee-ZHH-fig2}
\end{figure}

We numerically study the dependence of $A_i^{}$ on the parameters in our effective Lagrangian. The single Higgs couplings to vector bosons such as $HZZ$ may be precisely determined by measuring the polarization of the $Z$ boson in the process $e^+_{}e^-_{} \to ZH$~\cite{Rattazzi:1988ye, Barger:1993wt, Hagiwara:1993sw, Hagiwara:2000tk}. Therefore, we focus on the dependence on the parameters which cannot be accessed by single Higgs boson production processes. We choose as a benchmark point
\begin{align}
\delta_2^Z = \delta_2^A = \delta_2^{AA} = - \delta_3^Z = - \delta_3^A = - 0.05. 
\end{align}
The other parameters are set zero. The total cross section with this choice is in agreement with the SM value within $10\%$. In Figure~\ref{figure:ee-ZHH-fig1}, the observables $A_2^{}$, $A_5^{}$ and $A_6^{}$ are shown as deviations caused by adding non-zero parameters $\delta_9^{}$ (solid curve), $\delta_{10}^{}$ (dashed curve), $\delta_6^Z$ (dotted curve) and $\delta_7^{Z}$ (broken curve). In the left panel the parameters take positive values, and in the right panel the parameters take negative values. The results show that $A_{2,5,6}^{}$ depend little on $\delta_9^{}$. 
This indicates that ${\cal F}_{2,5,6}^{}$ have the similar dependences on $\delta_9^{}$ as ${\cal F}_1^{}$. 
The results also show that $A_{2,5,6}^{}$ depend largely on $\delta_{10}^{}$, $\delta_6^Z$ and $\delta_7^{Z}$. This indicates that the observables ${\cal F}_{2,5,6}^{}$ depend on these parameters in different ways from the total cross section. 
In the left panel of Figure~\ref{figure:ee-ZHH-fig2}, the CP-odd observables $A_8^{}$ and $A_9^{}$ are shown as deviations caused by adding non-zero CP-odd parameters $\tilde{\delta}_8^{Z}$ (solid curve) and $\tilde{\delta}_8^{Z}=\tilde{\delta}_8^{A}$ (dashed curve). The results show that $A_{8,9}^{}$ approach to zero as the CP-odd parameters become small, as expected. These observables are non-zero only when CP is violated. (Even if re-scattering effects exist, these observables are identically zero as long as CP is conserved.)\\

Due to the existence of the overall $\tau$ in the functions $F_{3}^{}$, $F_{4}^{}$ and $F_{7}^{}$, the corresponding observables $A_{3}^{}$, $A_{4}^{}$ and $A_{7}^{}$ can be suppressed. Longitudinally polarized $e^+_{}e^-_{}$ beams will be useful to study these 3 functions. We define observables as
\begin{align}
B_i^{} \equiv \frac{(1+P_-^{})(1-P_+^{}){\cal F}_i^{}(+) + (1-P_-^{})(1+P_+^{}){\cal F}_i^{}(-)}{(1+P_-^{})(1-P_+^{}){\cal F}_1^{}(+) + (1-P_-^{})(1+P_+^{}){\cal F}_1^{}(-)}, \label{eq:ObservableB}
\end{align}
where $P_-^{}$ ($-1 \le P_-^{} \le 1$) and $P_+^{}$ ($-1 \le P_+^{} \le 1$) denote the degrees of longitudinal polarization of the electron and the positron, respectively. We choose $(P_-^{}, P_+^{})=(-0.8, 0.3)$~\cite{Baer:2013cma}. In the right panel of Figure~\ref{figure:ee-ZHH-fig2}, the $\mathrm{CP\widetilde{T}}$-odd observables $A_7^{}$ and $B_7^{}$ are shown as deviations caused by adding non-zero imaginary parts in the parameters $\delta_6^Z$ (solid curve) and $\delta_{10}^{}$ (dashed curve)~\footnote{Re-scattering effects can be approximately included by allowing imaginary parts in the Higgs couplings~\cite{Hagiwara:1986vm}.}. For the results in this panel, we choose as a benchmark point
\begin{align}
\delta_2^Z = \delta_6^Z = - \delta_3^Z = - \delta_7^Z = \delta_{10}^{} = 0.1,
\end{align}
and the other parameters are set zero. The total cross section with this choice is in agreement with the SM value within $30\%$. The results show that $B_7^{} > A_7^{}$, i.e. the sensitivity to re-scattering effects can be significantly increased by means of longitudinally polarized $e^+_{}e^-_{}$ beams~\footnote{This benefit from polarized beams, however, becomes less clear when $HZ\gamma$, $H\gamma\gamma$ and/or $HHZ\gamma$ couplings are turned on, due to the interference between the diagrams exchanging the $Z$ boson and those exchanging the photon.}. These observables are non-zero only when re-scattering effects exist. (Even if CP is violated, these observables are identically zero unless re-scattering effects exist.) 
Note that the SM predictions in these $\mathrm{CP\widetilde{T}}$-odd observables are non-zero. The decay width in the $Z$ boson propagators contributes to these observables, since it indeed reflects the re-scattering effect of the light fermions in the propagating $Z$ boson. The contribution is, however, negligibly small as we can observe in our results (i.e. $A_7^{}$ and $B_7^{}$ approach to identically zero as the imaginary part of $\delta_6^Z$ or that of $\delta_{10}^{}$ becomes small.).

\section{Summary}\label{summary}

In this paper, we have derived the analytic expression for the differential cross section that in the most general case has the 9 non-zero functions $F_i^{}$ ($i=1,2,\cdots,9$), for the process $e^+_{}e^-_{}\to ZHH$. The functions $F_i^{}$ are the coefficients of the 9 angular terms, depend on the Higgs couplings, and can be experimentally measured as observables. 
We have re-derived the analytic forms of the observables which exist in the literature and have been widely used. 
We have found that all of these observables are directly related to $F_1^{}$, which is just one of our 9 functions. We have derived, for the first time, the analytic form of the $Z$ boson polar angle $\Theta$ distribution, which is related to 3 of our 9 functions: $F_1^{}$, $F_2^{}$ and $F_3^{}$. 
We have divided the 9 functions into 4 categories under CP and $\mathrm{CP\widetilde{T}}$: 4 even-even ($F_1^{}$, $F_2^{}$, $F_5^{}$, $F_6^{}$), 1 even-odd ($F_7^{}$), 2 odd-even ($F_8^{}$, $F_9^{}$) and 2 odd-odd ($F_3^{}$, $F_4^{}$). This result is summarized in Table~\ref{table:symproperty}. \\

We have introduced an effective Lagrangian for non-standard Higgs couplings to the Higgs boson itself, the $Z$ boson and the photon, and numerically studied the dependence of ${\cal F}_i^{}$ (this is obtained by integrating $F_i^{}$ over $Q^2_{}$ and $\cos{\xi}$; see eq.~(\ref{eq:IntegratedF})) on the parameters in the effective Lagrangian. 
For this purpose, we have formed new observables $A_i^{}$ and $B_i^{}$ (eqs.~(\ref{eq:ObservableA}) and (\ref{eq:ObservableB})) in terms of ${\cal F}_i^{}$. These observables are defined in such way that the differences between ${\cal F}_i^{}$ $(i=2,3,\cdots,9)$ and ${\cal F}_1^{}$ in the dependence on the Higgs couplings become apparent. Since ${\cal F}_1^{}$ is directly related to the total cross section (eq.~(\ref{eq:F1andCross})), by means of $A_i^{}$ and $B_i^{}$, we can learn whether ${\cal F}_i^{}$ $(i=2,3,\cdots,9)$ provide us different information about the Higgs couplings than the total cross section or not. 
We have found that the 3 observables ${\cal F}_{2,5,6}^{}$ have the similar dependences on the constant shift of the trilinear self-coupling of the Higgs boson (i.e. $\delta_9^{}$ in eq.~(\ref{eq:lagrangian})) as the total cross section, while they have quite different dependences on the other CP-even parameters (i.e. $\delta_{10}^{}$, $\delta_6^Z$ and $\delta_7^{Z}$) than the total cross section. This is shown in Figure~\ref{figure:ee-ZHH-fig1}. 
The 2 CP-odd observables ${\cal F}_{8,9}^{}$ and the $\mathrm{CP\widetilde{T}}$-odd observable ${\cal F}_{7}^{}$ clearly have advantages over the total cross section in determining CP-odd parameters and in observing re-scattering effects, respectively, since the total cross section is both CP-even and $\mathrm{CP\widetilde{T}}$-even. 
We have shown that the CP-odd observables ${\cal F}_{8,9}^{}$ directly measure CP violation, by showing that ${\cal F}_{8,9}^{}$ approach to identically zero as the CP-odd parameters (i.e. $\tilde{\delta}_8^{Z}$ and $\tilde{\delta}_8^{A}$) become small. This is shown in the left panel of Figure~\ref{figure:ee-ZHH-fig2}. 
Finally, we have shown that the use of longitudinally polarized $e^+_{}e^-_{}$ beams can enhance the ability of the $\mathrm{CP\widetilde{T}}$-odd observable ${\cal F}_{7}^{}$ which measures re-scattering effects. This is shown in the right panel of Figure~\ref{figure:ee-ZHH-fig2}.

\section*{Acknowledgments}

I sincerely appreciate the support from the Alexander von Humboldt Foundation.

\small

\bibliographystyle{JHEP}
\nocite{*}
\bibliography{zhh_bib}

\end{document}